# Giant Superelastic Piezoelectricity in Flexible Ferroelectric BaTiO$_3$ Membranes

*Hemaprabha Elangovan,[1,2,+] Maya Barzilay,[1,2,+] Sahar Seremi,[3,4,+] Noy Cohen,[1] Yizhe Jiang,[3,4] Lane W. Martin[3,4] and Yachin Ivry[1,2,*]*

[1]Department of Materials Science and Engineering, Technion – Israel Institute of Technology, Haifa 3200003, Israel.

[2]Solid State Institute, Technion – Israel Institute of Technology, Haifa 3200003, Israel.

[3]Department of Materials Science and Engineering, University of California, Berkeley, Berkeley, California 94720, USA.

[4]Materials Sciences Division, Lawrence Berkeley National Laboratory, Berkeley, California 94720, USA.

[*]E-mail: ivry@technion.ac.il

[+] These authors contributed equally to this work.

**Abstract**

Mechanical displacement in commonly used piezoelectric materials is typically restricted to linear or biaxial in nature and to a few percent of the material dimensions. Here, we show that free-standing $BaTiO_3$ membranes exhibit non-conventional electromechanical coupling. Under an external electric field, these superelastic membranes undergo controllable and reversible "sushi-rolling-like" 180° folding-unfolding cycles. This crease-free folding is mediated by charged ferroelectric domains, leading to a giant > 3.8 and 4.6 µm displacements for a 30-nm thick membrane at room temperature and 60 °C, respectively. Further increasing the electric field above the coercive value changes the fold curvature, hence augmenting the effective piezoresponse. Finally, it is found that the membranes fold with increasing temperature followed by complete immobility of the membrane above the Curie temperature, allowing us to model the ferroelectric-domain origin of the effect.

The electromechanical power conversion of piezoelectrics is the basis for a broad range of sensing, actuating, and communication technologies, including ultrasound imaging and cellular phones.[1–3] Recent interest in electromechanical energy harvesting[4,5] as well as in flexible electronics for wearable devices,[6,7] nano motors,[8] and medical applications[9–11] raises a need for flexible piezoelectric materials and devices. Modern applications of piezoelectrics hinge on thin films,[12–14] however, the substrate in such geometries is typically rigid, preventing the development of flexible devices. Flexible piezoelectric devices are therefore typically based on either nanowires[4] or on thin-film systems, but with substrates that have been designed especially for such applications.[15,16] Most piezoelectric applications rely on lead-based materials, which exhibit strong piezoelectric coefficients. Nevertheless, the toxicity of these materials is undesirable for environmental considerations, while it also disqualifies them for medical or wearable applications. Likewise, traditional thin-film geometries limit the electromechanical excitation modes. That is, usually, uniaxial electric field results in either parallel or perpendicular uniaxial or biaxial mechanical deformation (or vice versa). Nevertheless, the interest in flexible-electronic technologies raises a need for advanced electromechanical excitation modes, *e.g.*, for motorized devices, including microscale aerial vehicles.[17]

Substrate removal for piezoelectric films or membranes augments their functional properties,[18–21] mainly thanks to mechanically-induced ferroic-domain reorganization.[22] However, the preparation of completely stand-alone substrate-free films has remained a challenge. Lu *et al.*[23] demonstrated lately a general method to prepare oxide materials in the form of membranes, *i.e.*, continuous free-standing thin films with no substrate. More recently, Dong *et al.*[24] used this method to process $BaTiO_3$ membranes, which is a well-known lead-free piezoelectric and ferroelectric material. This work showed exceptional elastic and flexible properties that arise from the irregular ferroelectric domain dynamics in these membranes. Nevertheless, the piezoelectric properties of free-standing membranes have remained elusive. Here, we fabricated a flexible free-standing piezoelectric $BaTiO_3$ film. We show that under external electric fields, the membrane folds gradually and continuously by 180° similar to the rolling of a sushi roll with a bamboo makisu mat. The fold-unfold cycles are reversible and reproducible. We used this makisu-like piezoelectric effect to fold a 30-nm-thick membrane for a length that is greater than 4.6 μm. We demonstrate the temperature dependence of this unexpected electromechanical coupling mode, indicating that the effect is dominated by the polarization domains. Based on the thorough quantification of the electromechanical folding dynamics, we present a model that explains the domain origin of this unconventional piezoelectric effect.

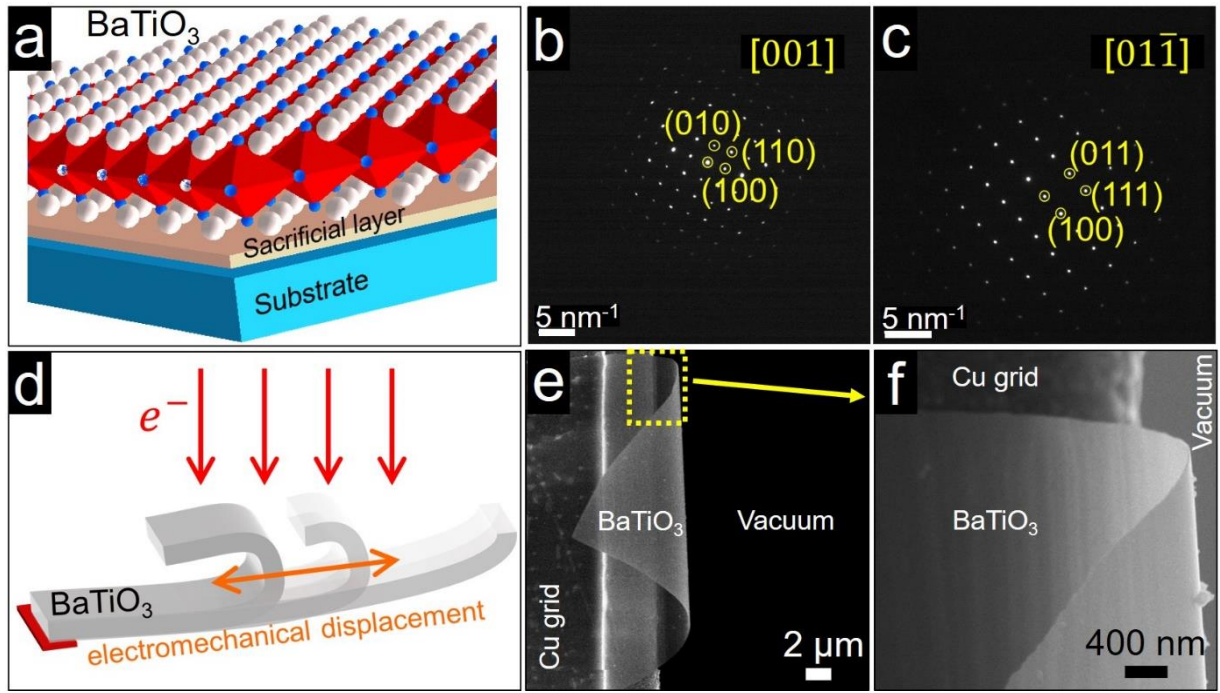

**Figure 1| Foldable single-crystal BaTiO₃ membrane.** (**a**) Schematic of the single-crystal (001) oriented BaTiO₃ membrane on the water-soluble sacrificial layer above the SrTiO₃ substrate, from which the membrane was extracted. Diffraction patterns of a 30-nm thick BaTiO₃ membrane from (**b**) [001] and (**c**) [01$\bar{1}$] zone axes showing the single-crystal tetragonal structure ($a = b = 3.95$ Å are the pseudo-cubic lattice constants). (**d**) Schematics of the folded membrane at the native state as well as its electron-mechanical rolling and unfolding response under an electron-beam induced field. (**e**) Scanning electron micrographs of the natively folded BaTiO₃ membrane and (**f**) a closer look micrograph from the fold region.

**Results and Discussions**

Free-standing single-crystal BaTiO₃ 30-nm thick membranes were prepared by pulsed-laser deposition when using a sacrificial layer that was dissolved in water (**Figure 1**a, see Methods for details as well as in Figure S1). The pseudo-cubic perovskite structure of the single crystal was confirmed by electron diffraction (Figures 1b-c). Membranes were folded in their native state, as illustrated both schematically (Figure 1d) and by scanning electron micrographs (Figures 1e-f). The samples were put on transmission electron microscope (TEM) carbon-free (100 μm) copper mesh and plasma cleaned. We used the TEM to exert electric field on the membrane *in situ* and image the resultant inverse piezoresponse.[25,26] Here, the electric field was adjusted by changing the electron-beam flux or dose (*D*) controllably, allowing quantitative measurements of the mechanical deformation. Details regarding the dose calibration and data acquisition are given in the Supplementary Information.

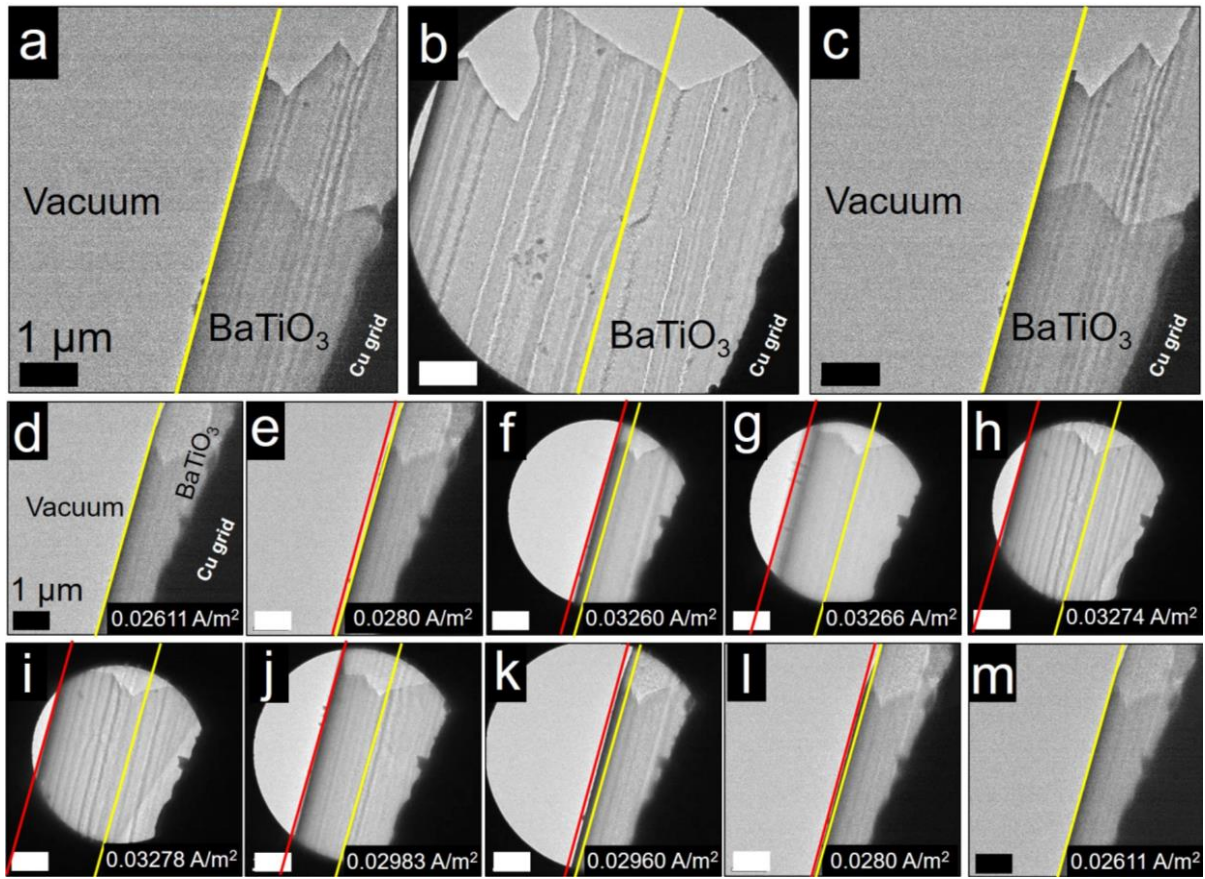

**Figure 2| Makisu-like folding cycle in BaTiO₃ membranes.** (**a**) TEM image of 30-nm thick BaTiO₃ membrane in its native state. The membrane is folded over itself at one edge and is tethered to the copper grid at its other edge (Sample #1). (**b**) At high dose values ($D$=0.0337 A/m$^2$), the membrane unfolds so that the structure extends in 3.75 µm. (**c**) When the electric field is reduced again, the membrane folds back to its initial state. (**d**-**h**) Representative TEM micrographs of an unfold process as a result of increasing dose. For low dose values, there is nearly no change in the fold (d-e). At intermediate dosage (f), above the threshold value $D_1$, the membrane unfolds, while maintaining constant radius of curvature at the fold, as in the case of sushi rolling. Finally, at high doses (above $D_2$), the sample unfolds abruptly with increasing electric field thanks to variation in the curvature at the fold (g-h), completing a three-step electromechanical response. (**i**-**m**) The electromechanical folding response as a function of decreasing dose showing the reversibility of the effect in the complete unfolding-folding cycle.

**Figure 2** shows the membrane dynamics under several representative dose values (the complete dynamics are given in Videos S1 and S2). Figure 2a shows that initially, at the native state, the membranes are folded. The bottom part of the membrane is tethered by the copper grid. At the other side, the edge of the membranes is folded over itself, *i.e.*, 180°, forming a top layer with an overlap of a length $L_0$. Upon increasing the electric field ($D = 0.0337$ A/m$^2$), the membrane unfolds and opens by > 3.8 µm, as seen in Figure 2b (actual unfold length is estimated as 4.2 µm, based on Video S1). This length is more than two orders of magnitude larger than the 30 nm thickness of the membrane. When the field is decreased, the membrane folds back to its initial native state (Figure 2c).

To characterize the electro-mechanical response over a complete unfold-fold period, we imaged and measured the mechanical displacement of the membrane under continuous variable electric-field conditions. The native membrane here is given in Figure 2d, where the top part is folded above the edge of the bottom part of the membrane. This fold remains constant even upon increasing the dose value, as long as the dose does not exceed a certain threshold value, $D_1$ (Figure 2e). When the dose is increased above $D_1$, the membrane unfolds continuously (Figures 2g-h). The unfolding is reminiscence of sushi rolling on a bamboo makisu mat, as seen in Figure 2. The electromechanical response of the makisu-like effect is reversible, with a small hysteresis between the unfolding and folding paths, while no crease is observed during or after the fold. The folding as a result of dose decrease is given in Figures 2i-m, completing the reversible unfolding-folding electromechanical response cycle.

To quantify the electromechanical response, we recorded the position of the fold for a representative experiment and plotted this value as a function of dose (**Figure 3**a). The complete piezoresponse of this cycle is given in Video S3. Figure 3a shows a clear distinction between the fold behavior at different dosage values. Below $D_1$, the membrane remains at its native state and there is no change in the folded length ($L_0$). At the intermediate dosage, between $D_1$ and $D_2$ (during the makisu-like rolling), the displacement of the fold front, or the length of the unfolding segment (designated with $L$ in Figure 3b), varies linearly with increasing dosage. The fold is thus described by: $L = L_0 + mD$, where $m$ is a constant and $L_0$, and $m$ are measurable. Here, the radius of curvature remains constant, so that the membrane rolls and the top and bottom parts of the membrane that are far from the fold look like sliding to opposite directions. At the second threshold dosage value, $D_2$ the membrane unfolds abruptly non-linearly with increasing dosage. Here, the rapid unfold occurs thanks to a significant sudden change in the curvature. Figure 3a shows also the reversible three-step effect, in which the membrane folds back towards its native state when the dosage is decreased. Schematic illustration of the membrane state is overlaid on Figure 3a for several dose values. The three-step electromechanical response is depicted schematically in Figure 3c (see also Video S3 for the experimental data that correspond to the three-step process).

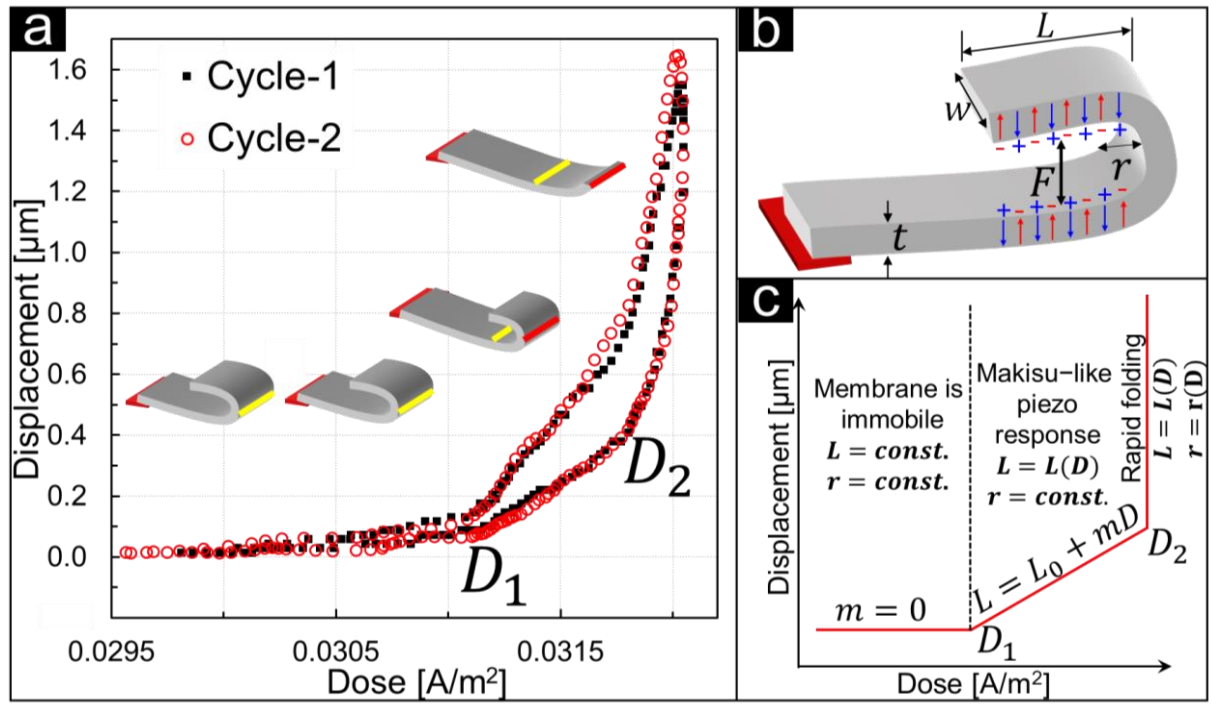

**Figure 3| Electro-mechanical cycles in BaTiO$_3$.** (**a**) Displacement of the membrane fold front *versus* dose (equivalent to electric field) for two cycles shows reversible and reproducible electro-mechanical response (data extracted from Sample #2). With increasing dosage, the fold remains unchanged in its native state until the value exceeds $D_1$. At intermediate dosage ($D_2 > D > D_1$), the membrane unfolds with linear dependence of the displacement on the dose similarly to the makisu mat motion during sushi rolling. When the dose reaches a threshold value ($D_2$), the membrane unfolds abruptly. Decreasing the dosage results is reversible piezoresponse, while repeating the electro-mechanical cycle demonstrates no apparent change in the reproducible piezoresponse. Additional data from three more samples are given in Video S4 to show the reproducibility of the effect. Schematic illustration of the membrane's electro-mechanical structure is given in (**b**), while illustration of the membrane fold for representative dose values that correspond to the micrographs in Figure 3 are overlaid in (a). (**c**) Simplified description of the piezoresponse cycle.

We then confirmed the reproducibility of this piezoresponse by performing a sequential unfold-fold cycle, as presented in Figure 3a (the complete cycles are given visually in Video S3). The observed reproducibility of the electro-mechanical cycles suggest that the fold does not involve creases. The universality of this effect was also confirmed by observing it in more than five different samples (Video S4 demonstrates such a typical measurement).

We then wanted to expose the origin of the giant electromechanical response. The change in dose may result in mechanical response due to, *e.g.*, irradiation effects, such as heating or damage. Such effects are examinable, because they are accompanied by a strong influence of the exposure time on the mechanical motion.[27] That is, the accumulated charge or power, not the flux or dose, are the driving excitations. Alternatively, if the effect is electromechanical, *i.e.*, due to piezoelectricity, there should be a dependence only on the electric field or dose value,[28,29] and not on the exposure time. We thus examined the dependence of the mechanical response on exposure time. We increased significantly the wait time whenever the dose was changed. Figure S2 shows that upon changing the dose, the sample unfolded only for a short time (2-3 sec), while increasing the exposure time beyond this time constant, for a given dose, resulted in no mechanical change in the fold. To further verify that the electromechanical effect is truly piezoelectric in nature, one should demonstrate dependence of the effect on the piezoelectric and perhaps also on the ferroelectric properties of the material. Piezoelectricity stems from the lack of inversion symmetry in the crystal structure. Although $BaTiO_3$ is polar and piezoelectric at room temperature, it undergoes a structural phase transformation at a Curie temperature (200 °C > $T_C$ > 100 °C).[30,31] Thus, above this Curie temperature ($T_C$), the crystal becomes centrosymmetric, eliminating piezoelectricity and ferroelectricity. To illustrate that the observed electromechanical response is piezoelectricity, one has therefore to confirm that the effect is absent above $T_C$.

**Figure 4** shows the fold evolution of the membrane as a function of *in situ* temperature variation (double-tilt heating holder, Gatan Inc. model 652), while complete unfold-fold cycles for various temperatures are given in <u>Video S5</u>. In Figure 3, the dose was held constant, so that the effect of temperature was isolated. The data clearly show that below 160 °C (Figure 4a-f), the length of the native membrane fold ($L_0$) increased with increasing temperature. Above $T$ = 150 °C, no electromechanical folding occurred (Figure 4g-i). Figure 4j shows quantification of the microscopy observations, where the temperature-dependent native fold ($L_{0_T}$) is plotted as a function of temperature. Below $T_C$, the piezoresponse cycles are similar to the room-temperature behavior of Figures 2-3. Because the initial fold ($L_{0T}$) increased with increasing temperature, complete open of the membrane gave rise to displacements larger than those recorded at room temperature, with a maximal displacement length of 4.6 µm. However, above $T_C$, the membrane became immobile and neither unfolded nor folded with dose variation. Therefore, not only do these results demonstrate that the electro-mechanical coupling is not heat driven, but they also confirm that the effect is based on the membrane's piezoelectric and ferroelectric properties. Note that the linear response at $D_2>D>D_1$ also indicates that the electromechanical response is piezoelectricity.

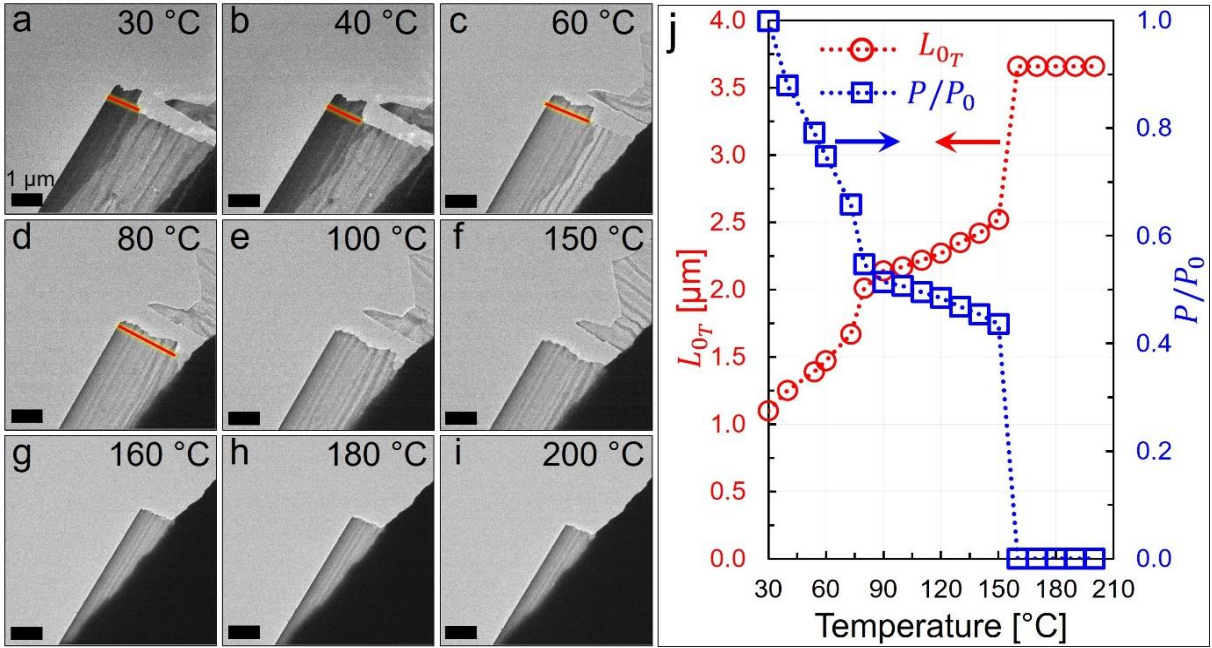

**Figure 4| Temperature dependence of the native fold in a BaTiO$_3$ membrane.** (**a**-**f**) TEM micrographs of the native membrane ($D \ll D_1$) with *in situ* temperature variation shows an increase of the folded segment ($L_0$) with increasing temperature, for temperatures up to 150 °C (*i.e.*, the Curie temperature). (**g**-**i**) Above 150 °C, the native fold did not change with temperature. Dose variation for the different temperatures ([Video S5](Video S5)) show that the membrane undergoes electro-mechanical cycles up to 150 °C and above this temperature, the membrane remains immobile. (**k**) Temperature dependence of the native fold length ($L_{0_T}$) as well as of the corresponding calculated polarization (see Equation 2).

A "sushi-rolling-like" motion does not usually occur in continuous structures in nature, and it requires the motion of discrete jointed elements, similar to the motion of the discrete bamboo sticks in a makisu mat or to the motion of inch-worm motors. Moreover, the temperature dependence of the fold (Figure 4j) is reminiscence of the dependence of polarization on temperature in ferroelectrics. Hence, one can assume that the folding-unfolding piezoresponse emerges from the discrete polarization distribution, which follows the domain structure along the membrane. This assumption receives strong support from the recent work of Dong *et al.,*[24] in which the effect of polarized ferroelectric domains on the elastic properties of such membranes was found to be crucial.

Here, we propose that the top and bottom parts of the fold are being held by attractive electric forces that compensate for the stress, which is developed by the curved fold. The electromechanical motion is possible due to ferroelectric domains with a finite width. Following Roytburd's model,[32] a 30-nm thick membrane has 28-nm domain periodicity. That is, domains from the top and bottom parts of the membrane form surface charge with opposite polarity, forming an effective plate capacitor (see SI for details of the model derivation). The two charged surfaces attract each other, thanks to a force between opposite domains with width $w$, length $l$ and surface charge $\sigma$, which is given by: $F_{p\pm} = \frac{wl}{\epsilon_0 \epsilon} \sigma^2$. Here, $\epsilon_0$ is the vacuum permittivity and $\epsilon$ is the dielectric constant of the medium between the two parts of the membrane.

The total force between all domains should sum up to the mechanical force that is required to bend a membrane with Young's modulus[33] $Y$ and a thickness $t$ in a radius of curvature $r$. Thus, we obtain a direct relationship between the surface charge and the length of the fold:

$$\sigma = \frac{1}{L}\sqrt{\frac{\epsilon\epsilon_0 Y t^3}{6r}} \quad (1)$$

That is, increasing the dose, which gives rise to a decrease in $L$ (see Figure 3c), stems from an increase in the surface charge. Presumably, the electromechanical effect was related to direct electron charging; however, the absence of a dependence of the electromechanical effect on time exposure suggests that the effect is not due to direct-charge injection. We may therefore attribute the surface charge, $\sigma$, broadly to the out-of-plane domain polarization, $P$.

At the native state ($D<D_1$), the surface charge corresponds to $P$.[34] Applying a dose is equivalent to exerting an external electric field on the membrane thanks to the high dielectric constant of ferroelectrics.[28,29,35] High electric field and hence high doses are expected to allow polarization switching, while electric-field gradients may also induce similar effects due to flexoelectricity.[36,37] Typically, for dc fields, there is no time dependence (beyond a certain Debye time constant) of the domain distribution for a given electric field.[38] When the dose increases, the polarization changes to reduce the global electromechanical free energy of the system. This change occurs gradually at first ($D_2>D>D_1$). At a certain threshold value, ($D = D_2$), the polarization changes abruptly, suggesting that $D_2$ is equivalent to the global coercive electric field (expected to be around 2 V for our membrane).[24] This analysis complies with existing models that show that domains in free-standing $BaTiO_3$ membranes are rather easily switchable.[24] Hence, we can deduce the dependence of polarization on the observable fold length for doses smaller than $D_2$ as:

$$\frac{P}{P_0} = \frac{L_0}{L} \quad (2)$$

Note that for at $D_2>D>D_1$, $L$ changes linearly with the dose (Figures 3a and 3c), so that the dependence of $P$ on $D$ can be expressed directly. Moreover, Equation 1 is valid for the polarization dependence on $L$ also at higher-dose values, above $D_2$. However, in this regime, the radius of curvature changes, adding an unknown free parameter to the system, and the exact fitting of the experimental results becomes more challenging. Likewise, we should note that for such high doses, the effect of direct charging by the beam – not necessarily of the membrane, but also of surface adsorbents[39] – may become meaningful, also effectively affecting $\epsilon$.

Equation 2 allows us to analyze the temperature-dependence experiment (Figure 4). Here, the initial fold at the native state $\left(L_{0_T}\right)$ increases with increasing temperature. Substituting $L$ with $L_{0_T}$ in Equation 2 allows us to plot the temperature dependence of the polarization in the membrane (Figure 4k). That is, the decrease in polarization with temperature reduces the surface charge and hence increases $L_{0_T}$; whereas above $T_C$, there is no polarization and $L_{0_T}$ remains constant. Of course, for temperatures higher than $T_C$, Equation 2 is not valid. Yet, surface charge and charged adsorbents may still give rise to a fold, following Equation 1. Here, there are no domains that change with dose and hence there should not be any folding effect. These conclusions are in agreement with the experimental observations in Video S5. Similarly, Video S6 shows temperature reversibility of the electromechanical folding.

**Conclusions**

As a final remark, we would like to summarize our observations in the light of possible future applications and research directions. The electromechanical response as well as the unexpected functionality of the piezoelectric-ferroelectric freestanding membranes presented here is applicable for a broad range of devices and technologies, including at the micrometer and nanometer scale. These devices are completely different from piezoelectric systems that are based on bulk materials, or on thin films. The reversible crease-free and a "sushi-rolling-like" folding are exceptional mechanical motions that do not usually exist in continuous materials in nature and invite further investigation. The demonstrated reversible and reproducible behavior is promising for robust motors and actuators with high-precision controllability. Likewise, the temperature dependence of the makisu-like piezoresponse can be utilized for temperature-driven actuators, sensors, and motors. In addition, the present demonstration of contact-less electromechanical actuation enables contact-less piezoelectric devices that do not require constant voltage contacts and can be operated remotely, which is important, *e.g.*, in environments that do not allow electric wiring. Likewise, the domain-driven giant electro-mechanical response presented here experimentally and theoretically deserves further insight into the fundamental mechanism, while seeking additional systems that utilize this effect may result in technologically and scientifically advantageous materials.

**Experimental Methods**

**Thin-Film Deposition**

To produce the free-standing films, we first optimized the growth of $BaTiO_3/SrTiO_3$ (001) heterostructures using pulsed-laser deposition with a KrF excimer laser (248 nm, Compex, Coherent), in an on-axis geometry. X-ray diffraction studies (Figure S1a) reveal the ability to produce high-quality, fully epitaxial films of $BaTiO_3$. In turn, to produce the free-standing films, 30-nm $BaTiO_3$/20-nm $La_{0.7}Sr_{0.3}MnO_3/SrTiO_3$ (001) heterostructures were grown *via* pulsed-laser deposition. The $La_{0.7}Sr_{0.3}MnO_3$ films that served as a sacrificial etching layer for the subsequent release of the film from the substrate were grown directly on the $SrTiO_3$ (001) single-crystal substrates (Crystec, GmbH) from a ceramic $La_{0.7}Sr_{0.3}MnO_3$ target, at a heater temperature of 650 °C in a dynamic oxygen pressure of 200 mTorr at a laser repetition rate of 3 Hz and a laser fluence of 1.03 J $cm^{-2}$. The $BaTiO_3$ films were grown from a $BaTiO_3$ ceramic target at a heater temperature of 600 °C in a dynamic oxygen pressure of 40 mTorr at a 2 Hz laser repetition rate and a laser fluence of 1.14 J $cm^{-2}$. The substrates were adhered to the heater with silver paint (Ted Pella, Inc.). Following growth, the heterostructures were cooled to room temperature at a rate of 10 °C $min.^{-1}$ in a static oxygen pressure of ~700 Torr. Details regarding the $BaTiO_3$ film deposition and the membrane processing are in Figure S1.

**Film Release and Transfer to a TEM Grid**

After growth, the $BaTiO_3$ films were released from the substrate by selective etching of the sacrificial $La_{0.7}Sr_{0.3}MnO_3$ layer. The etching process requires that the heterostructures be placed in an aqueous solution of 8 mg KI + 10 ml HCl + 100 ml $H_2O$; this solution results in a high etch rate for $La_{0.7}Sr_{0.3}MnO_3$, while having negligible impact on the $BaTiO_3$ films (*i.e.*, etch rate of $La_{0.7}Sr_{0.3}MnO_3$ is much greater than the etch rate of $BaTiO_3$). After selectively etching the $La_{0.7}Sr_{0.3}MnO_3$, the released $BaTiO_3$ thin films were observed to sit on the substrate. The unclamped thin films were then removed from the substrate by slowly dipping the substrate in deionized water. When done correctly, the released films float on the water surface (Figure S1b). A nichrome wire loop (Ted Pella, Inc.) was then used to "catch" the freestanding $BaTiO_3$ films from the water surface (Figure S1c). The films were held in place by a thin layer of water which was formed inside the metal loop. The released films were then moved to a secondary substrate (*i.e.*, a 100 mesh carbon-free copper TEM grid, Ted Pella, Inc.). A piece of filter paper was placed underneath the TEM grid, so that when the loop came close to the grid the water inside the loop is absorbed by the paper, thus leaving the film on the TEM grid (Figure S1d). Tapping mode (AC Air) atomic force microscopy studies of the $BaTiO_3$ surfaces before and after release show no changes in topography and RMS roughness, which remained around 400 pm.

**Dose Calibration and Data Acquisition**

The Condenser-2 (C2) lens strength of the TEM determines the electron dose. Therefore, by changing the strength of the C2 lens (using intensity knob), the electron dose has been increased or decreased.

The TEM software presents the measured dose value, which is obtained by integrating the intensity of the beam falling on the fluorescent screen. During imaging and video recording, electron beam that goes through the sample falls on the camera and not on the fluorescent screen. Therefore, only the C2 lens strength will be indicated in the software during imaging. Hence, dose value correspondence to the C2 lens strength is captured, by allowing the beam to fall on the fluorescent screen. Later, the mechanical motion has been recorded along with the C2 lens strength when allowing the beam to fall on the camera. The C2 lens strength and the dose value has been interpolated from these two experiments.

**Structural Characterization**

Structural studies were performed using high-resolution X-ray diffraction (Panalytical, X'pert3 MRD) 9 kW diffractometer. A Cu $k_\alpha$ rotating-anode source at 45 kV was used, with a 200 mA tube current. TEM studies were performed using a Technai T20 microscope equipped with $LaB_6$ thermionic emission gun operated at 200 keV. The TEM samples were plasma cleaned under $Ar+O_2$ atmosphere for 10 sec prior inserting to the TEM. The Scanning Electron Microscope (SEM) micrographs were acquired using Zeiss Ultra-Plus FEG-SEM operated at 4 keV (Figure 1e) and 20 keV (Figure 1f) acceleration voltage.

**Acknowledgments**

The Technion team acknowledges support from the Zuckerman STEM Leadership Program and the Technion Russel Barry Nanoscience Institute, as well as from the Israel Science

Foundation (ISF) grant #1602/17. We would also like to thank Dr. Yaron Kauffman and Mr. Michael Kalina for technical support. S.S. acknowledges support from the National Science Foundation under grant DMR-1708615. Y.J. acknowledges support from the U.S. Department of Energy, Office of Science, Office of Basic Energy Sciences, under Award Number DE-SC-0012375 for the development of free-standing ferroelectric materials. L.W.M. acknowledges support from the U.S. Department of Energy, Office of Science, Office of Basic Energy Sciences, Materials Sciences and Engineering Division under Contract No. DE-AC02-05-CH11231 (Materials Project program KC23MP) for the development of functional ferroic materials.

# Supplementary Information

## SI-Table of contents

## BaTiO$_3$ release process and XRD profile

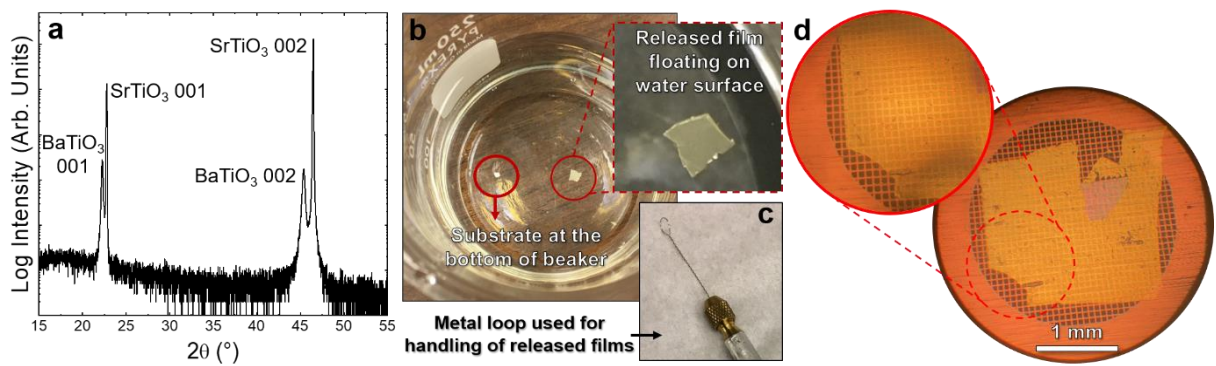

**Figure S1 | High-quality BaTiO$_3$ films and the release process.** (**a**) X-ray diffraction profile reveals the production of fully epitaxial BaTiO$_3$ films. Optical micrographs are given to help explain the preparation process. (**b**) Carefully placing the substrate and released film in water allows for the film to be removed from the substrate and float on the liquid surface. (**c**) A metal loop is used to pick-and-place released films without damaging them. (**d**) The resulting freestanding films are of relatively large area and exhibit very few cracks and asperities.

**Time-dependence study**

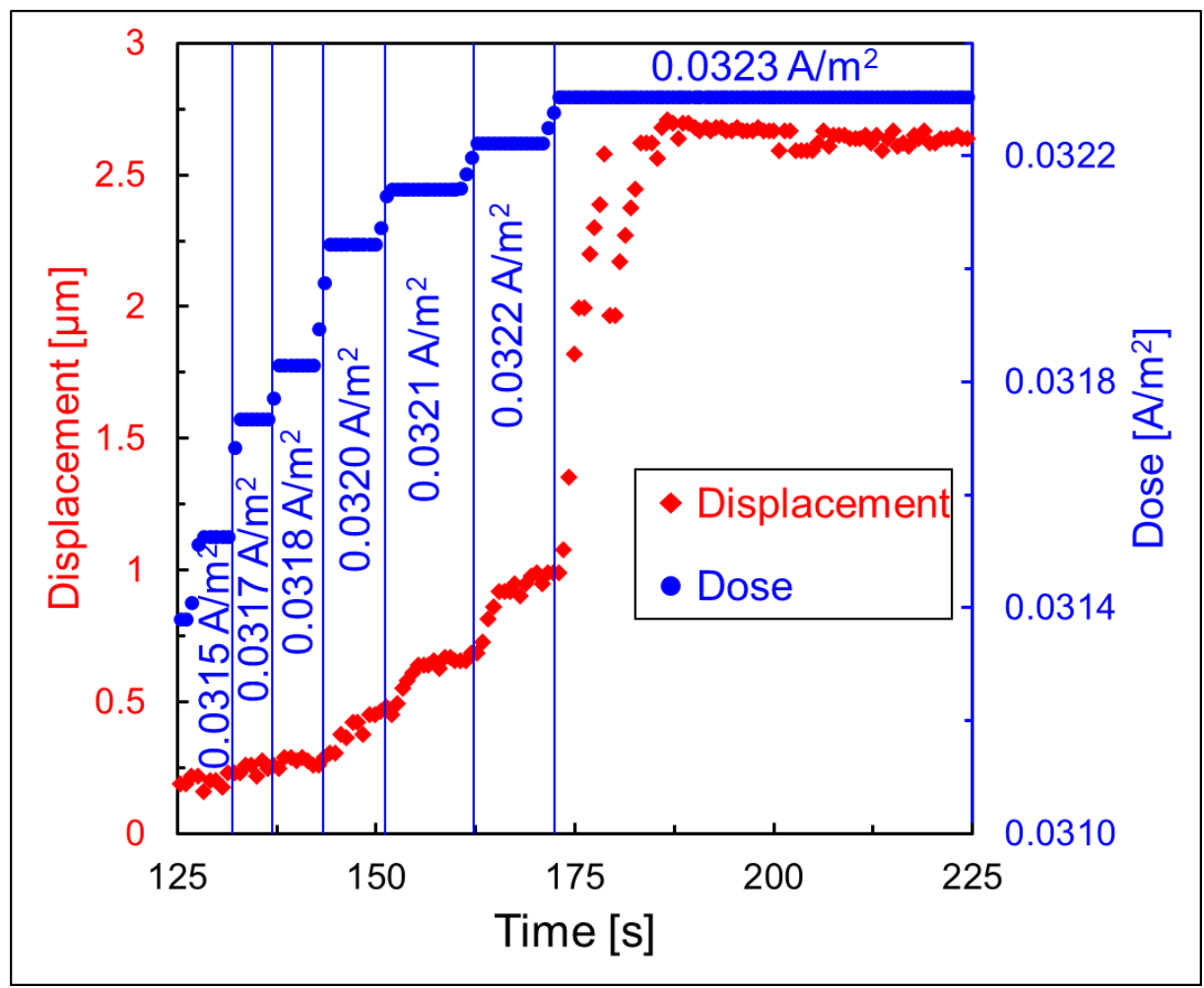

**Figure S2. Electromechanical fold during slowly changing dosage**. Displacement of the fold with the simultaneous values of electron-beam dose for increasing dosage. Here, the membrane unfolds immediately whenever the dose changes (with a time constant of 2-3 seconds). Maintaining the dosage constant does not induce any electromechanical motion, even at high doses. Hence, the electromechanical folding-unfolding cycles are a result of change in the dosage and neither of the absolute value of the dose, nor of the irradiation exposure time. Note that because the electromechanical motion depends on the rate of increase of dose, if a sample is exposed to a rapid change in dosage, the membrane displacement may fluctuate. That is, the sample transitions fast back and forth from folding to unfolding with a significant displacement.

**Model**

Assuming that the folded region of the membrane corresponds to a charged parallel-plate capacitor, the force between the top and bottom layer of the membrane is: $F = qE$. Here, $E$ is the electric field that is induced by the surface charge $\sigma$, given by $E = \sigma/\epsilon\epsilon_0$ ($\epsilon$ are the dielectric constant of the medium between the plates and $\epsilon_0$ is vacuum permittivity), where $q$ is the total charge and $A$ is the plate area. The shorter side of the membrane, which has length $L$ and width $w$ (as well as thickness $t$) defines the capacitor size, *i.e.*, the total electric surface charge is:

$$q = \int \sigma dA = \sigma L w \quad (S1)$$

and the force becomes:

$$F = \frac{wL}{\epsilon\epsilon_0} \sigma^2 \quad (S2)$$

From elemental beam-theory view, the attractive electric force between the top and bottom sides of the folded membrane creates a torque on the fold: $M_F = F \cdot \frac{L}{2}$. In equilibrium, this value equals to the bending torque that is required to fold a beam with bending stiffness $B$ to a radius of curvature $r$: $M_B = \frac{B}{r}$, while we assumed that the curvature satisfies: $\kappa = 1/r$.

Substituting Equation S2 in $M_F$ and equating $M_F$ and $M_B$, we obtain:

$$\left(\frac{wL}{\epsilon\epsilon_0} \sigma^2\right) \cdot \left(\frac{L}{2}\right) = \frac{B}{r} \quad (S3)$$

The bending stiffness is the product of Young's modulus ($Y$) of the membrane and the moment of inertia of the folded region ($I = wt^3/12$): $B = YI$, so that we can rewrite Equation S3:

$$\left(\frac{w}{2\epsilon\epsilon_0} \sigma^2 L^2\right) = \frac{Y \cdot \frac{wt^3}{12}}{r} \quad (S4)$$

And we can now finally extract the relationship between the length of the folded region and the surface charge (*i.e.*, Equation 1):

$$L = \frac{1}{\sigma}\sqrt{\frac{\epsilon\epsilon_0 Y t^3}{6r}} \qquad (S5)$$

In the ferroelectric state, the surface charge that raises the attractive force between the folded sides of the membrane stems from the collective organization of dipole moments in the form of polarization domains. The attraction arises from oppositely charge surfaces, *i.e.,* domains with opposite polarity (see Figure 3b). That is, the membrane arranges to minimize the net charge within the folded region. This minimization is first done by motion of the membranes that allows maximal proximity between opposite charges (*i.e.,* native state and makisu-like folding). For higher doses, the electric field exceeds the electric field and the domain switch, giving rise to rapid unfold/fold motion with varying $r$. The dielectric constant of the medium between the folded membrane originates from surface adsorbents. For high electric field (high doses), these adsorbents may be charged, so that $\sigma$ exists also above the Curie temperature. Note that the length of the membrane fold ($L_0$) is crucial in determining the dose required for folding ($D_1$), as well as the displacement increase between $D_1$ and $D_2$ (see in Figure 3c).

The surface charge is due to both intrinsic depolarization and external adsorbents, which are likely to be due to the polarization field as well. Adsorbents have a non-zero net charge, especially in the vacuum conditions of the TEM, while the vacuum may affect also the intrinsic surface charge.[39] Note that the domains may not be perfectly aligned to each other. Rather, it is sufficient to have an overlap between surfaces from both parts of the fold with opposite charge. These opposite charges give rise to the attraction between the two parts of the fold and they change with dosage, so that the attraction changes controllably. Thus, integration of local attraction between individual domain pairs from the top and bottom parts of the membrane give rise to the total attraction, even though the net global polarization might be zero. An upper barrier for our calculations is obtained when the domains are aligned.

Note that the exact time of the domain formation is currently beyond the scope of the current experimental results and model. Following Roytburd's model,[32] we expect domain to exist prior removal from the substrate. Likewise, following the model from Dong *et al.,*[24] the domains exist also in the flat membrane after the removal from the substrate.

The folded membranes are of a typical fold length ($L$) of 1 to 3 µm. As the temperature increases, $L$ increases and the corresponding experimental relationship is given in Figure 4j. This relationship helps us extract the surface charge, by using Equation 1. In this equation, $\epsilon_0$ is the vacuum permittivity and $\epsilon$ is the dielectric constant of the medium between the two parts of the membrane. A possible medium due to the surface adsorbents is considered with a value of 100.[39] Typical Young's modulus of the BaTiO$_3$ is considered (65 GPa).[33] Measured thickness of the membrane was 30 nm and we assumed here a radius of curvature a few nanometers for the folding-unfolding observations. Based on the data from Dong *et al.*,[24] we assume that the radius of curvature is equals roughly to the films thickness (note that this radius is smaller than the values imaged with SEM, which are of a different geometry than the membranes in the folding-unfolding experiments). Although neighboring domains have opposite polarity, the attraction between the top and bottom parts of the membrane is between individual domains. For calculating the attraction forces, we integrated on the absolute value of the surface charge for the entire fold length (1 µm). Substituting these values into Equation 1 we obtained that the entire the surface charge integration, $\sigma$ or the polarization $P$ (Equation 1) is 0.16 C m$^{-2}$, which is comparable to the bulk polarization.[31]

During unfolding, for the lower dosage values, the domains are not likely to undergo complete transformation, at least not a complete transformation. Rather, the electron beam modifies the surface charge and not the domains themselves. This can be due to either direct charging of the BaTiO$_3$ surface or due to adsorbent charging and mobility. Domains may stop being homogeneous due to the beam,[35] *i.e.*, the beam may induce random nucleation sites of switched domains, so that the net polarization of a given domain is decreased.

For large dose values (>$D_2$), the domains switch, hence the large increase in displacement. The domains of the folded membrane have a preferable orientation (see model in Ref. [24]) and a driving force that acts to restore it. Hence, although domains are switchable with an external field, the switched state is stable only when the external field is applied. Upon removal or decrease of this field, the domains relax to the stable state (relaxation to preferable domain orientation state as a function of strain[16]). The existence of this mechanism is supported also by the time-dependent experiments under a nearly constant electron dosage (Figure S2). Here, the displacement is persistent only when the dose is applied, while upon releasing the dose, the membrane returns to its native state.

Finally, note that the domains may not be perfectly aligned to each other. Rather, it is sufficient to have an overlap between surfaces from both parts of the fold with opposite charge. These opposite charges give rise to the attraction between the two parts of the fold and they change with dosage, so that the attraction changes controllably. Yet, an upper barrier is obtained when the domains are aligned.

**Videos**

[Video S1](): Folding-unfolding cycle under variable dosage captured for Sample #1 (shown in Figure 2a-c). Running time is six times faster than in the originally captured data.

[Video S2](): An electromechanical unfolding-folding cycle under variable dosage captured for Sample #4 (shown in Figure 2d-m). Running time is four times faster than in the originally captured data.

[Video S3](): Electromechanical unfolding-folding cycles (two consecutive cycles) under variable dosage captured for Sample #2 (shown in Figure 3). The membrane motion and the value of the $D_1$ and $D_2$ are consistent in the two cycles, demonstrating reversibility and reproducibility. Running time is six times faster than the originally captured data.

[Video S4](): Folding-unfolding electromechanical cycles for two different samples (Sample #3 and Sample #5), showing the universality of the electromechanical folding-unfolding behavior upon application of an electric field on $BaTiO_3$ membrane. Running time is four times faster than the originally captured data.

[Video S5](): Folding-unfolding cycles for sample #1 captured at different temperatures, below and above $T_C$. The electromechanical motion is observed for temperatures less than $T_C$. Above $T_C$, the membrane remains immobile, showing that the electromechanical motion is polarization dependent. Running time is four times faster than the originally captured data except for the experiment at 200 °C, which is played two times faster than the originally captured data.

[Video S6](): Temperature reversibility of electromechanical folding-unfolding: a folding-unfolding electromechanical cycle for sample #6 captured at 30 °C after cooled down from 200 °C, showing similar behavior as before heating (*e.g.*, Videos S1-S5). Note that a complete heating-cooling cycle requires very long exposure times that to introduce extrinsic effects (*e.g.*, strong adhesion of immobile adsorbents,[39] chemical destruction[27]). Running time is ten times faster than the originally captured data.